\documentclass[aps,superscriptaddress,showpacs,twocolumn]{revtex4}
\usepackage{amssymb}

\usepackage{amsmath}
\usepackage{graphicx}
 
\begin{document}

\title{Phase ambiguity of the threshold amplitude in $pp \rightarrow pp\pi^0$}
\author{G. Ramachandran}
\affiliation{G. V. K. Academy, Jayanagar, Bangalore - 560082, India}
\author{G. Padmanabha}
\affiliation{G. V. K. Academy, Jayanagar, Bangalore - 560082, India}
\affiliation{Department of Physics, Bangalore University, Bangalore - 560056, India}
\author{Sujith Thomas}
\affiliation{G. V. K. Academy, Jayanagar, Bangalore - 560082, India}
\affiliation{K. S. Institute of Technology, Bangalore - 560056, India}

\pacs{13.75.Cs, 13.75.Gx, 21.3Cb, 25.10+s, 24.70+s, 25.40Ve}

\begin{abstract}
\begin{center}
{\bf Abstract}
\end{center}
Measurements of spin observables in $pp \rightarrow \vec{p}\vec{p}\pi^0$ are suggested to remove the phase ambiguity of the threshold amplitude. The suggested measurements complement the IUCF data on $\vec{p}\vec{p}\rightarrow pp\pi^0$ to completely determine all the twelve partial wave amplitudes, taken into consideration by Mayer et.al. \cite{may} and Deepak, Haidenbauer and Hanhart \cite{de4}.
\end{abstract}
\maketitle

\begin{section}{Introduction}

Meson production in $NN$ collisions has continued to excite considerable interest ~\cite{mac,mos,fal,han},
since total cross-section measurements ~\cite{totcr} for $pp\rightarrow pp\pi^0$ in the early
$1990$'s were found to be more than a factor of five larger than the then available theoretical
predictions~\cite{thpre}. To bridge the gap between experiment and theory, several mechanisms
like, exchange of heavy mesons, two pion exchange, off shell extrapolation of the vertex form
factor, final state interactions, contributions due to $\Delta$ resonance and of low lying 
nucleon resonances, were proposed. Hanhart et al. ~\cite{hanhart} in $2000$ have observed:
`` As far as microscopic calculations of the reaction $NN\rightarrow NN\pi$ are concerned, one
has to concede that theory is definitely lagging behind the experimental sector.....Further more 
they take into account only the lowest partial wave(s). Therefore, it is not possible to confront
these models with the wealth of experimental data available nowadays with differential cross-sections and with spin dependent observables." The Julich model, on the other hand, takes into consideration higher partial waves as well.

In contrast to elastic $NN$ scattering where channel spin is conserved, the $pp\rightarrow pp\pi^0$
transition at threshold to the final $Ss$ state is a triplet to singlet. Next in order are the 
transitions to $Ps$ states which are singlet to triplet. As the energy is increased transitions
to $Pp$ states are also expected to contribute, which are, however, triplet to triplet. Pionic $d$-wave effects were reported~\cite{dwave} even at a beam energy of $310$ MeV. Measurements 
upto $425$ MeV have also been reported ~\cite{bilger},  where evidence for $Ds$ state
was seen even at $310$ MeV. Advances in storage ring technology~\cite{mey}  led to detailed
experimental studies, including measurements of 
spin observables employing polarized beams of protons on polarized proton targets. Of the
two existing models ~\cite{models,mae} which include higher partial waves, the Julich meson
exchange model~\cite{hanhart,models} was thoroughly confronted with these data. The model was comparatively more
successful with the less complete data on ${\vec p}{\vec p}\rightarrow d\pi^+$~\cite{prez}
and ${\vec p}{\vec p}\rightarrow pn\pi^+$~\cite{dae}, but failed to provide an overall
satisfactory reproduction of the complete set of polarization observables in the case of 
${\vec p}{\vec p}\rightarrow pp\pi^0$~\cite{may}. In this context, a model independent 
approach~\cite{gr1,gr2}, was developed, using irreducible tensor techniques~\cite{gr3}. The reaction is characterized, in this formalism, by irreductible tensor amplitudes $M^\lambda_\mu(s_f,s_i)$ of rank $\lambda = |s_f-s_i|,..,(s_f+s_i)$, where $s_i,s_f$ denote the initial and final spin states of the two protons. Each of these amplitudes is expressible in terms of partial wave amplitudes $M^j_{l(l_fs_f)j_f;l_is_i}$, which are functions of the c.m. energy $E$ and invariant mass $W$ of the two proton system in the final state. The relative orbital angular momentum between the two protons in the initial and final states are denoted by $l_i$ and $l_f$ respectively and $l$ denotes the pion orbital angular momentum in the c.m frame. The threshold amplitude $M^0_{0{00}0;11}$ contributes to $M^1_0(0,1)$, and an empirical estimate of the integrated $|M^1_0(0,1)|^2$ was presented  in~\cite{gr1}, based on the then existing data~\cite{totcr}. The same approach was employed subsequently to analyze~\cite{de1} the IUCF data on ${\vec p}{\vec p}\rightarrow pp\pi^0$~\cite{may} immediately after its publication. The sixteen partial waves listed by Mayer et al.~\cite{may}, covered the $Ss, Ps, Pp, Sd$ and
$Ds$ channels. Here, the capital letters denote $l_f$ while the lower case indicate $l$. In~\cite{de4}, the same set of partial waves were listed, of which, the last four, covering $Sd$ and $Ds$ were ignored following~\cite{may}. Since, the final spin-singlet and spin-triplet states do not mix in any of the spin observables measured in~\cite{may}, the $Ss$ amplitude and the larger of the $Ps$ amplitude were both chosen to be real in~\cite{de4}. This implies that the phase of the $Ss$ amplitude remained ambiguous but chosen to be zero with respect to the larger $Ps$ amplitude.
The comparison of the empirically extracted amplitudes with the Julich model 
predictions revealed that $i)$ the $\Delta$ contributions are important, $ii)$ the model 
deviated very strongly in the case of ${^3}P_1\rightarrow {^3}P_0p$ and to a lesser extent in 
${^3}F_3\rightarrow {^3}P_2p$ which `` will guide the search for the possible shortcomings."~\cite{de4}
  
The purpose of the present paper is to extend the model independent theoretical discussion
to the spin polarization of the protons in the final state and to examine how the additional experimental
measurements regarding the final spin state can be used to determine empirically, the strengths of all these amplitudes and the ambiguous relative phase of the threshold $Ss$ amplitude with respect to the eleven near threshold $Ps$ and $Pp$ amplitudes, considered in \cite{may,de1,de4}
\end{section}
\begin{section}{Theoretical Formalism}
We consider the reaction $pp\rightarrow pp \pi^{0}$ at c.m. energy $E$ and initial c.m. momentum ${\bf p_{i}}=p_{i}\bf \hat p_{i}$ which may be chosen to be along the $z$-axis. Let ${\bf q}=q\bf \hat q=-(\bf {p_{1}+p_{2}})$ denote the pion momentum in c.m. frame and let ${\bf {p_{f}}}=p_{f}\bf{\hat p_{f}}=\frac{1}{2}(\bf {p_{1}-p_{2}})$ in terms of the c.m. momenta $\bf p_1$ and $\bf p_2$ of the two protons in the final state.

Following~\cite{gr1}, we write the matrix $M$ in spin space for the reaction $pp \rightarrow pp \pi^0$ in the form
\begin{equation}
M = \sum_{s_i,s_f=0}^1 \sum_{\lambda=|s_i-s_f|}^{s_i+s_f} (S^\lambda (s_f,s_i)\cdot M^\lambda(s_f,s_i)),
\label{mm}
\end{equation}
where $s_i$ and $s_f$ denote the initial and final channel spins respectively. The
irreducible tensor operators $S^\lambda_\mu (s_f,s_i)$ of rank $\lambda$ with $\mu$ taking values $\mu = \lambda,\lambda-1,....,-\lambda$
are defined in ~\cite{gr3}. The irreducible tensor amplitudes $M^\lambda_\mu(s_f,s_i)$ in $(1)$
are expressible as 
\begin{eqnarray}
M^\lambda_\mu(s_f,s_i) &=& \sum_{{\cal L},{j}}\,W(l_is_iL_fs_f;j\lambda) \,Z(s_f,s_i,{\cal L},j)\nonumber \\  & & \times\,A^\lambda_\mu({\cal L}),
\label{m}
\end{eqnarray}
where \begin{equation}
A^\lambda_\mu({\cal L}) = ((Y_{l_f}({\bf \hat p}_f)\otimes Y_{l}({\bf \hat q}))^{L_f}\otimes Y_{l_i}({\bf \hat p}_i))^{\lambda}_\mu\,,
\end{equation}
and the symbol ${\cal L}$ is used to collectively denote ${\cal L}\equiv \{l_f, l,L_f,l_i\}$. It may be noted that $(-1)^{l_f+l+l_i} = -1$ due to parity conservation. The complex numbers $Z(s_f,s_i,{\cal L},j)$ are given by
\begin{eqnarray}
Z(s_f,s_i,{\cal L},j) & = & \frac{[L_f][j]^{2}}{[s_f]}\,(-1)^{j-{s_i}+1}\,\sum_{j_f} \,[j_f]\,\nonumber\\
& & \times\,W(s_fl_fjl;j_fL_f)M^j_{l(l_fs_f)j_f;l_is_i}
\label{z}
\end{eqnarray}
in terms of the sixteen partial wave reaction amplitudes
\begin{eqnarray}
M^j_{l(l_fs_f)j_f;l_is_i} & = & F\,\langle(l(l_fs_f)j_f)j||T||(l_is_i)j\rangle,\;\;\,
\label{M}
\end{eqnarray}
proportional to the reduced on-energy-shell $T$ matrix elements $\langle (l(l_fs_f)j_f)j||T||(l_is_i)j\rangle $ for the reaction. The purely kinematical factor 
\begin{equation}
F=(-i)^{l_i-l-l_f}4(2\pi)^{1/2}\sqrt{{W\,\omega(E-\omega)qp_f}/{p_i}}\,,
\end{equation}
is introduced explicitly in (\ref{M}), so that the dependence on $E$ and $W$ is seen to be completely taken care of by the $M^j_{l(l_fs_f)j_f;l_is_i}$. They are identical to the amplitudes denoted as $T$ in \cite{de4}.We may, following~\cite{may,de4}, neglect the last four amplitudes which are $Sd$ and $Ds$ and consider the first twelve amplitudes, which are, for simplicity, enumerated as $f_1,....,f_{12}$ in Table \ref{Tone}.

The unpolarized double differential cross section may now be written as
\begin{equation}
{\frac{d^2\sigma_0}{dW\,d\Omega_f\,d\Omega}} = {\textstyle \frac {1}{4}}Tr[M\,M^\dagger],
\label{N}
\end{equation} 
where $M^\dagger$ denotes the hermitian conjugate of $M$ given by \ref{mm}). The invariant mass $W$ of the two protons in the final state is given by
\begin{equation} 
W = \sqrt{({E^2 + m^2_\pi-2 E \omega})}\,,
\end{equation}
where $m_\pi$ denotes the pion mass and $\omega$ denotes the c.m. energy of pion. It may be noted that 
\begin{equation}
\frac{d^2\sigma_0}{d^3p_f\,d\Omega} = \frac{W}{p_f}\,\frac{d^2\sigma_0}{dWd\Omega_fd\Omega},
\end{equation}
  
\begin{table}
\caption{\label{Tone}List of the partial wave amplitudes for the reaction ${\vec p}{\vec p}\rightarrow pp\pi^0$}
\begin{ruledtabular}
\begin{tabular}{cccc}
\hline
Initial $pp$  &Type & Final $pp\pi^0$ & Partial Wave \\
state & & state & Amplitudes \\
\hline
${^3}P_0$ & $Ss$ & ${^1}S_0,s$ & $M^0_{0(00)0;11}= f_1$\\
\hline
${^1}S_0$ & $Ps$ & ${^3}P_0,s$ & $M^0_{0(11)0;00}= f_2$\\
${^1}D_2$ & &${^3}P_2,s$ & $M^2_{0(11)2;20}= f_3$\\
\hline
${^3}P_0$ &$Pp$ &${^3}P_1,p$ & $M^0_{1(11)1;11}= f_4$\\
${^3}P_2$ & &${^3}P_1,p$ & $M^2_{1(11)1;11}= f_5$\\
${^3}P_2$ & &${^3}P_2,p$ & $M^2_{1(11)2;11}= f_6$\\
${^3}F_2$ & &${^3}P_1,p$ & $M^2_{1(11)1;31}= f_7$\\
${^3}F_2$ & &${^3}P_2,p$ & $M^2_{1(11)2;31}= f_8$\\
${^3}P_1$ & &${^3}P_0,p$ & $M^1_{1(11)0;11}= f_9$\\
${^3}P_1$ & &${^3}P_1,p$ & $M^1_{1(11)1;11}= f_{10}$\\
${^3}P_1$ & &${^3}P_2,p$ & $M^1_{1(11)2;11}= f_{11}$\\
${^3}F_3$ & &${^3}P_2,p$ & $M^3_{1(11)2;31}= f_{12}$\\
\hline
\end{tabular}
\end{ruledtabular}
\end{table}
 It is worth noting that the threshold $Ss$ amplitude $f_1$ alone contributes to 
\begin{equation}
M_{\mu}^{1}(0,1)=\frac{1}{4\,\sqrt{3}\,\pi}f_1Y_{1\mu}(\hat{\bf p_i})\,,
\label{m1mu}
\end{equation}
which is spherically symmetric both w.r.t $\hat{\bf p_{f}}$ as well as $\hat{\bf q}$ in the final state, while all the other irreducible tensor amplitudes are independent of $f_1$. 
\end{section}
\begin{section} {Final state polarization with initially unpolarized protons.}
     If the colliding protons are unpolarized, the spin density matrix $\rho^f$ characterizing the two protons in the final state is given by
\begin{equation}
\rho^f = {\textstyle \frac{1}{4}}\, M\,M^\dagger\,,
\label{rhof}
\end{equation}
so that (\ref{N}) is identical with ${Tr[\rho^f]}$.

The final spin state is completely determined through measurements of the polarizations 
\begin{equation}
{\bf P}_i = \frac{Tr[\boldsymbol \sigma_{i} \, \rho^f]}{Tr[\rho^f]}\, , i=1,2
\end{equation}
of the two protons and their spin-correlations 
\begin{equation}
C_{\alpha \beta} = \frac {Tr [\boldsymbol \sigma_{1\alpha}\,\boldsymbol \sigma_{2\beta}\,\rho^{f}]}{Tr[\rho^f]}\,,\, \alpha,\beta = x,y,z.
\label{corr}
\end{equation}

All these spin observables may elegantly be calculated by considering
\begin{equation}
P^k_\mu(s'_f,s_f) = {Tr [S^k_\mu(s'_f,s_f)\,\rho^f]},
\label{pkmu1}
\end{equation}
where $S^k_\mu({s'_f,s_f})$ are given in terms of the Pauli spin matrices ${\boldsymbol \sigma}_1$ and ${\boldsymbol \sigma}_2$ of the two protons in the final state through 
\begin{eqnarray}
S^0_0(0,0)&=& {\textstyle \frac{1}{4}} (1 - {\boldsymbol \sigma}_1\cdot{\boldsymbol \sigma}_2)
\label{s00}\\
S^0_0(1,1)&=& {\textstyle \frac{1}{4}} (3 + {\boldsymbol \sigma}_1\cdot{\boldsymbol \sigma}_2)\\ 
S^1_\mu(1,1) &=& {\textstyle \frac{\sqrt 3}{2\sqrt 2}}({\boldsymbol \sigma}_1+{\boldsymbol \sigma}_2)^1_\mu 
\label{s1mu}\\
S^2_\mu(1,1) &=& {\textstyle \frac{\sqrt 3}{2}} ({{\boldsymbol \sigma}_1\otimes{\boldsymbol \sigma}_2})^2_\mu \\
S^1_\mu(0,1) &=& {\textstyle \frac{1}{2\sqrt2}}({{\boldsymbol \sigma}_1\otimes{\boldsymbol \sigma}_2})^1_\mu -{\textstyle \frac{1}{4}} ({\boldsymbol \sigma}_1-{\boldsymbol \sigma}_2)^1_\mu
\label{s101} \\ 
S^1_\mu(1,0) &=& {\textstyle \frac{\sqrt 3}{2\sqrt2}}({{\boldsymbol \sigma}_1\otimes{\boldsymbol \sigma}_2})^1_\mu +{\textstyle \frac{\sqrt 3}{4}} ({\boldsymbol \sigma}_1-{\boldsymbol \sigma}_2)^1_\mu\,. 
\label{ss} 
\end{eqnarray}
Thus, the double differential cross section is given by 
\begin{equation}
{\frac{d^2\sigma_0}{dW\,d\Omega_f\,d\Omega}} = Tr[\rho^f] = P^0_0(0,0) + P^0_0(1,1)\,,
\label{cs}
\end{equation}
in terms of the double differential cross sections $P^0_0(0,0)$ leading to the final singlet state and $P^0_0(1,1)$ leading to the final triplet state of the two protons. If we use the notations $({\bf P_i})_\mu$ to denote the spherical components, i.e.,
\begin{equation}
({\bf P_i})_0=P_{iz}; ({\bf P_i})_{\pm 1}=\mp{\textstyle \frac{1}{\sqrt2}}\,(P_{ix}\pm P_{iy})\,,
\label{eq:}
\end{equation}
it follows from (\ref{s101}) and (\ref{ss}) that 
\begin{equation}
P^1_\mu(1,0)-\sqrt{3}P^1_\mu(0,1)=\frac{\sqrt{3}}{2}\,Tr[\rho^f]\,({\bf P}_1-{\bf P}_2)_\mu, 
\label{pmu2}
\end{equation}
where as it follows from (\ref{s1mu}) that
\begin{eqnarray}
P^1_\mu(1,1)&=&\frac{\sqrt{3}}{2\sqrt{2}} Tr[\rho^f]\,({\bf P}_1+{\bf P}_2)_\mu,
\end{eqnarray}
which together determine ${\bf P}_1$ and ${\bf P}_2$ individually. Finally, the spin correlations $C_{\alpha \beta}$ defined in (\ref{corr}) may like wise be related to (\ref{pkmu1}) using
\begin{equation}
P^0_0(1,1)-3P^0_0(0,0)=Tr[({\boldsymbol \sigma}_1 \cdot{\boldsymbol \sigma}_2)\,\rho^f]\,,
\end{equation}
\begin{eqnarray}
P^1_\mu(1,0) + \sqrt{3} P^1_\mu(0,1)&=&\frac{\sqrt{3}i}{2}\,Tr[\rho^f\,(\boldsymbol \sigma_1\times\boldsymbol \sigma_2)]_\mu,
\label{pmu1}
\end{eqnarray}
\begin{equation}
P^2_\mu(1,1)=\frac{\sqrt{3}}{2} Tr[\rho^f\,(\boldsymbol \sigma_1\otimes\boldsymbol \sigma_2)^2_\mu]\,.
\end{equation}

Using the known properties~\cite{gr3} of the spin operators $S^\lambda_\mu$ and standard Racah techniques, we may obtain a master formula for all the final state spin observables, which is given by
\begin{eqnarray}
P^k_\mu(s'_f,s_f) &=& {\textstyle \frac{1}{4}} \sum_{s_i,\lambda,\lambda'} (-1)^{s_f-s_i} [s_f]\,[s'_f]^2\,[\lambda][\lambda']\nonumber \\
& & \times\, W(s'_f\lambda's_f\lambda;s_i k)\nonumber \\
& & \times\,(M{^\lambda}(s_f,s_i)\otimes M^{^\dagger\lambda'} (s'_f,s_i))^k_\mu\,,
\label{pkmu}
\end{eqnarray}\\
where $M^{^\dagger\lambda}_{\mu} (s_f,s_i)$ are defined in terms of the complex conjugates $M^{\lambda}_{\mu} (s_f,s_i)^*$ of $M^{\lambda}_{\mu} (s_f,s_i)$ given by (\ref{m}) through
\begin{eqnarray}
M^{^\dagger\lambda}_{\mu} (s_f,s_i) &=& (-1)^\mu M^{\lambda}_{-\mu} (s_f,s_i)^*,
\end{eqnarray}
Noting once again that $(-1)^{l_f+l+l_i}=-1$, due to parity conservation, we may express
\begin{eqnarray}
M^{^\dagger\lambda}_{\mu} (s_f,s_i)&=&(-1)^{1-\lambda}\,\sum_{{\cal L}} W(l_is_iL_fs_f;j\lambda)\,\nonumber \\
& &\times Z^*(s_f,s_i,j,{\cal L})\,A^\lambda_\mu({\cal L})\,,
\label{mmud}
\end{eqnarray}
where $Z^*(s_f,s_i,j,{\cal L})$ denote the complex conjugates of $Z(s_f,s_i,j,{\cal L})$ given by (\ref{z}).
\end{section}
\begin{section} {RELATIVE PHASE OF THE THRESHOLD AMPLITUDE}
We may now take advantage of the fact that $M^1_0(0,1)$ given by (\ref{m1mu}) is spherically symmetric with respect to ${\hat{\bf p}_f}$ and ${\hat{\bf q}}$ and involves only the threshold amplitude $f_1$. Moreover, $M^{\lambda}_{\mu}(1,1)$ are independent of $f_1$ and depend only on the $Pp$ amplitudes $f_4,......,f_{12}$. Therefore, we focus attention on (\ref{pmu1}) and (\ref{pmu2}) which involve
\begin{eqnarray}
(M^{\lambda}(1,1)\otimes M^{^\dagger1}(0,1))^{1}_{\mu}&=&\frac{1}{4\,\sqrt{3}\,\pi}\sum_{{{\cal L},j}}W(l_i1L_f1;j\lambda)\nonumber\\
& &\times Z(1,1,j,{\cal L})f_{1}^{*}\nonumber\\
& &\times (A^{\lambda}({\cal L})\otimes Y_{1}(\hat{\bf p_i}))_{\mu}^{1}\,,
\label{mlmdl}
\end{eqnarray}
\begin{eqnarray}
(M^{1}(0,1)\otimes M^{^\dagger\lambda}(1,1))^{1}_{\mu}&=&\frac{-1}{4\,\sqrt{3}\,\pi}\sum_{{{\cal L},j}}W(l_i1L_f1;j\lambda)\nonumber\\
& &\times  Z^{*}(1,1,j,{\cal L})f_{1}\nonumber\\
& &\times (A^{\lambda}({\cal L})\otimes Y_{1}(\hat{\bf p_i}))_{\mu}^{1}\,,
\label{mmdll}
\end{eqnarray}
Expressing
\begin{eqnarray}
(A^{\lambda}({\cal L})\otimes Y_{1}(\hat{\bf p_i}))_{\mu}^{1}&=&\frac{\sqrt{3}}{4\pi}\sum_{L_i}W(L_fl_i11;\lambda L_i)\nonumber\\
& &\times [\lambda][L_i][l_i]C(l_i1L_i,000)\nonumber\\
& &\times A^1_{\mu}(l_flL_fL_i) \,,
\end{eqnarray}
and carrying out the summation over $\cal L$ and $j$, we obtain
\begin{eqnarray}
P^1_\mu(1,0)&=&{f^*_1}[F_1A^1_{\mu}(1110)\nonumber\\
& &+F_2A^1_{\mu}(1112)+F_3A^1_{\mu}(1122)]\,,
\end{eqnarray}
\begin{eqnarray}
P^1_\mu(0,1)&=&\frac{-1}{\sqrt{3}}\,{f_1}[F_1^*A^1_{\mu}(1110)\nonumber\\
& &+F_2^*A^1_{\mu}(1112)+F_3^*A^1_{\mu}(1122)]\,,
\end{eqnarray}
where $F_{i}, i=1,2,3$ are well-defined linear combinations of the $Pp$ amplitudes given by 
\begin{eqnarray}
F_1&=&\frac{1}{32{\pi}^{3/2}}\,[f_4-\frac{5}{6}f_5+\frac{5}{2\sqrt{3}}f_6+\frac{1}{3\sqrt{3}}f_9-\frac{1}{6}f_{10}\nonumber\\
& &-\frac{\sqrt{5}}{6\sqrt{3}}f_{11}]\,,
\end{eqnarray}
\begin{eqnarray}
F_2&=&\frac{1}{32\sqrt{2}{\pi}^{3/2}}\,[f_5-\sqrt{3}f_6+\sqrt{\frac{3}{2}}f_7+\frac{3}{\sqrt{2}}f_8]\,,\;\;\;\;
\end{eqnarray}
\begin{eqnarray}
F_3&=&\frac{-1}{32\sqrt{2}{\pi}^{3/2}}\,[\sqrt{3}f_5+f_6+\sqrt{7}f_7+\sqrt{\frac{7}{3}}f_8]\,,\;\;\;\;
\end{eqnarray}
Since the $Pp$ amplitudes have been determined both in magnitude and relative phase  w.r.t $f_2$ in \cite{de4}, we may express
\begin{eqnarray}
F_{\alpha}=\left|F_{\alpha}\right|e^{i\Delta_{\alpha}}\,,\alpha=1,2,3
\end{eqnarray}
and treat $\left|F_{\alpha}\right|$ and $\Delta_{\alpha}$ as known. In \cite{de4}, $f_2$ was assumed to be real. Since the relative phase between $f_1$ and $f_2$ could  not be ascertained from the measurements of Meyer et al. \cite{may}, $f_1$ was also assumed to be real, although only one of the amplitudes can be taken as real. Therefore, we choose $f_2$ to be real and express $f_1$ as
\begin{eqnarray}
f_1=\left|f_1\right|e^{i\delta_1}\,,
\end{eqnarray}
This leads to
\begin{eqnarray}
P^1_\mu(1,0)-\sqrt{3}P^1_\mu(0,1)=2\sum_{\alpha=1}^{3}R_{\alpha}cos({\Delta_{\alpha}}-{\delta_1})A_{\mu}^{1}({\alpha}),\;\;\,
\label{p1mu+}
\end{eqnarray}
\begin{eqnarray}
P^1_\mu(1,0)+\sqrt{3}P^1_\mu(0,1)=2i\sum_{\alpha=1}^{3}R_{\alpha}sin({\Delta_{\alpha}}-{\delta_1})A_{\mu}^{1}({\alpha}),\;\;\,
\label{p1mu-}
\end{eqnarray}
where $R_{\alpha}=\left|F_\alpha\right|\left|f_1\right|$ and $A_{\mu}^{1}({\alpha})$ for ${\alpha}=1,2,3$ denote $A_{\mu}^{1}(1110), A_{\mu}^{1}(1112), A_{\mu}^{1}(1122)$ respectively. 

It is seen from (\ref{rhof}) that measuring the double differential crosssection (\ref{N}) yields $Tr[\rho^{f}]$. Measurements of $({\bf P}_1-{\bf P}_2)_{\mu}$ given by (\ref{pmu2}) then leads to empirical determination of (\ref{p1mu+}), while measurements of spin correlations $C_{xy}-C_{yx}, C_{yz}-C_{zy}, C_{zx}-C_{xz}$ where $C_{\alpha\beta}$ are given by (\ref{corr}) lead to empirical determination of (\ref{p1mu-}) using (\ref{pmu1}).

Thus, we find that it is possible to determine empirically the relative phase $\delta_{1}$ of $f_1$, without any trigonometric ambiguities, since $R_{\alpha}$ and $\Delta_{\alpha}$ are known from \cite{de4}. We therefore advocate measurement of these $pp$ spin observables in the final state, employing simply an unpolarized beam and unpolarized target initially, to complement the spin observables measured by Meyer et al.\cite{may}, so that the amplitudes $f_1,f_2,.....f_{12}$ may be determined empirically without any phase ambiguity.
\end{section}

\end{document}